\documentclass[prl,twocolumn,superscriptaddress,showpacs]{revtex4}

\usepackage{graphicx}
\usepackage{amssymb}
\usepackage{multirow}

\begin{document}

\title{Influence of copper on the electronic properties of amorphous 
chalcogenides}

\author{S. I. Simdyankin}
\email{sis24@cam.ac.uk}
\affiliation{Department of Chemistry, University of Cambridge,
Lensfield Road, Cambridge CB2 1EW, United Kingdom}

\author{M. Elstner}
\affiliation{German Cancer Research Center, Dept. Molecular Biophysics,
             D-69120 Heidelberg, Germany}
\affiliation{Fachbereich 6~---~Theoretische Physik, Universit\"at Paderborn,
             Warburger Stra{\ss}e 100, D-33098, Paderborn, Germany}

\author{T. A. Niehaus} 
\affiliation{German Cancer Research Center, Dept. Molecular Biophysics,
             D-69120 Heidelberg, Germany}
\affiliation{Fachbereich 6~---~Theoretische Physik, Universit\"at Paderborn,
             Warburger Stra{\ss}e 100, D-33098, Paderborn, Germany}

\author{Th. Frauenheim}
\affiliation{Fachbereich 6~---~Theoretische Physik, Universit\"at Paderborn,
             Warburger Stra{\ss}e 100, D-33098, Paderborn, Germany}

\author{S. R. Elliott}
\affiliation{Department of Chemistry, University of Cambridge,
Lensfield Road, Cambridge CB2 1EW, United Kingdom}

\date{\today}

\begin{abstract}

We have studied the influence of alloying copper with amorphous arsenic
sulfide on the electronic properties of this material.
In our computer-generated models, copper is found in two-fold near-linear and
four-fold square-planar configurations, which apparently correspond to Cu(I)
and Cu(II) oxidation states.
The number of overcoordinated atoms, both arsenic and sulfur, grows with
increasing concentration of copper.
Overcoordinated sulfur is found in trigonal planar configuration, and
overcoordinated (four-fold) arsenic is in tetrahedral configuration.
Addition of copper suppresses the localization of lone-pair electrons on
chalcogen atoms, and localized states at the top of the valence band are due
to Cu $3d$ orbitals.
Evidently, these additional Cu states, which are positioned at the same
energies as the states due to [As$_4$]$^-$-[S$_3$]$^+$ pairs, are responsible
for masking photodarkening in Cu chalcogenides.

\end{abstract}

%78.66.Jg, 71.55.Jv, 73.61.Jc

\maketitle

%\section{Introduction}

There has been a significant amount of experimental research on the role of
copper in amorphous chalcogenide alloys.
It is generally believed that copper is found in tetrahedral configuration and
is bonded exclusively to chalcogen atoms ~\cite{Liang_1974,Saleh_1989}.
However, as noted in Ref.~\cite{Liang_1974}, the interpretation of
experimental radial distributions of ternary systems is necessarily ambiguous.
More recent experimental results \cite{Adriaenssens_2000} indicate that copper
also bonds to arsenic as well as to chalcogen atoms.
It was observed \cite{Liu_1987} that photodarkening (red shift of the optical
absorption edge under illumination) disappears in
Cu$_x$(As$_{0.4}$Z$_{0.6}$)$_{1-x}$ alloys, starting from $x=1$\% for Z=S and
$x=5$\% for Z=Se.
Two possible interpretations of this observation, given in
Ref. \cite{Liu_1990}, are either that the photodarkening is masked by some
electronic states due to Cu atoms, or the addition of copper interferes with
correlations of lone-pair orbitals on a scale greater than that of a
nearest-neighbor distance.
Alternatively, the Cu-related states at the band edges may be efficient
non-radiative centers which provide an alternative channel to that resulting
in photostructural bond rearrangement of the chalcogenide network.
The observation that photoinduced phenomena, other than photodarkening, remain
intact after the addition of copper \cite{Bolle_1998} favors the first
interpretation.

It is a natural next step in the investigation of these materials to obtain
the first detailed description of their electronic and structural properties
by means of computer simulation.
This is the topic of this paper.

%\section{Methodology}

Following Refs.~\cite{Simdyankin_PRL_2005,Simdyankin_2004}, we have employed a
density-functional-based tight-binding (DFTB) method
\cite{Porezag_95,Frauenheim_2002}, which has allowed us to create realistic
models of arsenic sulfide (As$_2$S$_3$).
The DFTB method allows one to improve upon the standard tight-binding
approximation by including a so-called self-consistent charge (SCC) correction
\cite{Elstner_1998}, derived from density-functional theory (DFT), to the
total energy.
The procedure for generating the tabulated data sets for pairwise interatomic
interactions is outlined in Ref.~\cite{Niehaus_01}.
The S-S \cite{Niehaus_01}, As-As and As-S \cite{Niehaus_unpub}, and Cu-Cu,
Cu-S and Cu-As \cite{Elstner_unpub} tables, and information on their creation,
are available from the authors.
The following electron orbitals were included in the basis set: 3$s$, 3$p$,
and 3$d$ for S atoms; 4$s$, 4$p$, and 4$d$ for As atoms; and 3$d$, 4$s$, and
4$p$ for Cu atoms.
Using the $d$ orbitals, in addition to the valence $s$ and $p$ orbitals on the
S and As atoms, has proved to be crucial to describe hypervalent compounds
\cite{Niehaus_01} and defect centers
\cite{Simdyankin_PRL_2005,Simdyankin_Sicily_2004}.

%\section{Results}

In the description of defects and impurities in crystalline materials, an
ideal crystal is the obvious reference system.
The situation is more complex with amorphous solids, where no atomic
configuration is unique.
It is, however, still possible to use a stoichiometric (all-heteropolar)
structural model as a useful reference for a binary amorphous compound.
In this way, any deviation from the reference-system topology and chemical
order can be considered as a defect, in direct analogy with the case of a
crystal.
Here, we use a stoichiometric 200-atom model of amorphous (a-)As$_2$S$_3$ from
Ref.~\cite{Simdyankin_2004}, reoptimized with a richer basis set, with
$d$ orbitals added to the As atoms.
This model will be referred to as model~0, in the following.

The other models, described in the following, have been prepared by adding one
(model 1, 1.6\% Cu), two (model 2, 3.2\% Cu), and four (model 3, 6.3\% Cu)
copper atoms to a 60-atom (24 As and 36 S) all-heteropolar model of
a-As$_2$S$_3$ \cite{Simdyankin_PRL_2005}.
As we performed spin-unpolarized calculations, the identity of one S atom in
model~1 was changed to As, in order to keep the total number of electrons in
the model to be even.
The mass densities (volumes per atom) of models 1, 2, and~3 are 3.282, 3.304,
and 3.226 g/cm$^3$ (25.371, 24.962, and 25.791 {\AA}$^3$/atom), respectively.
Keeping in mind possible local-density fluctuations in bulk amorphous alloys,
by slightly varying the density of the models, we verified that the results
are not particularly sensitive to this parameter for the above changes in the
volume per atom.
The configurations of the models used for subsequent analysis were prepared by
following the melt-and-cool schedule described in
Ref.~\cite{Simdyankin_2004}.
Models 1, 2, and 3 correspond to energy-minimized snapshots of $T=300$K
configurations, and model 3a corresponds to an energy-minimized snapshot of a
$T=700$K configuration, where all four copper atoms were two-fold coordinated.

\begin{figure}[t]
\centerline{\hfill (a)\includegraphics[width=3.6cm]{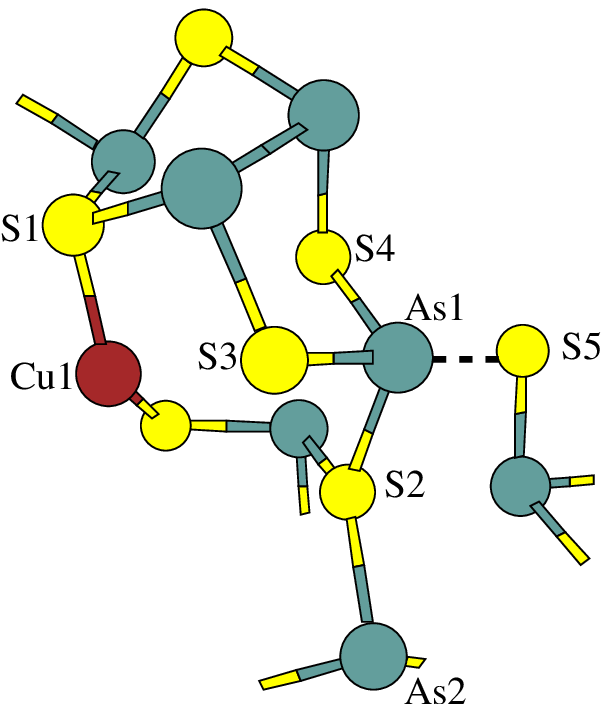} \hfill
(b)\includegraphics[width=3.2cm]{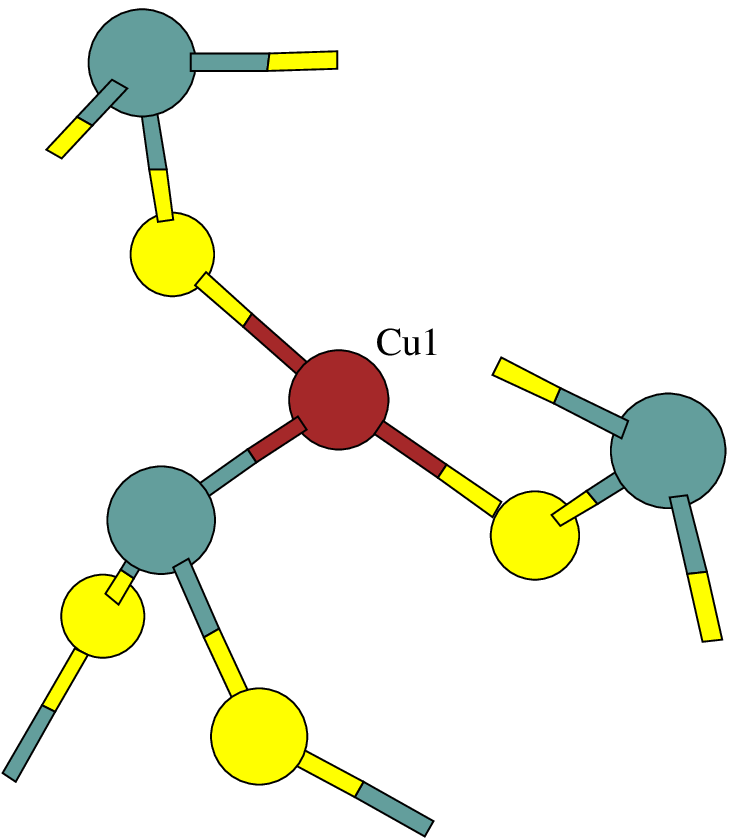} \hfill}
\centerline{\hfill (c)\includegraphics[width=4cm]{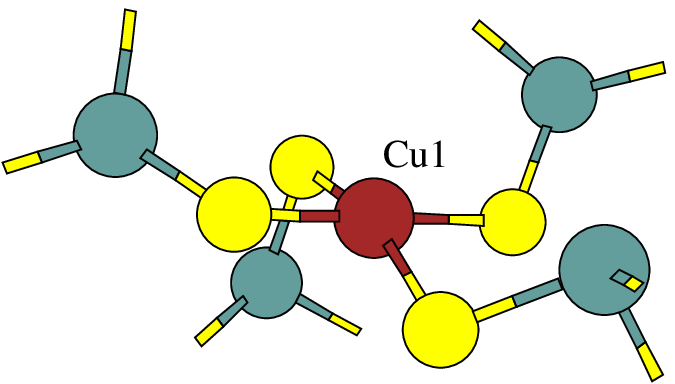} \hfill
(d)\includegraphics[width=4cm]{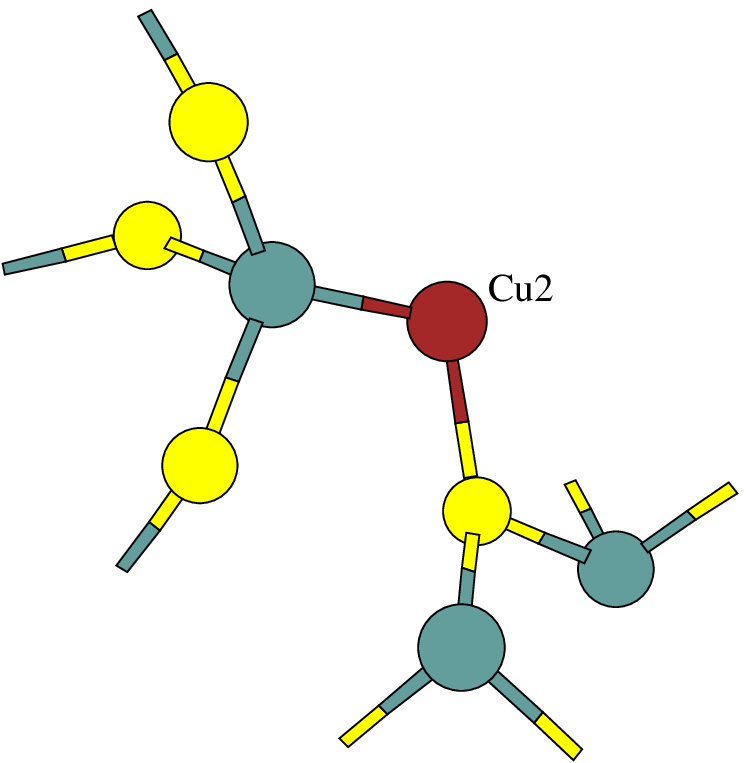} \hfill} 
\caption{Fragments of models 1 (a), 3 (b), and 2 (c,d) containing copper
atoms.  The dangling bonds show where the displayed configurations connect to
the rest (not shown) of the network. Note that the fragments in (c) and (d)
are from the same model, but they do not contain common atoms and are shown
with independent and arbitrary orientations. The numbering of Cu atoms
corresponds to Table~\ref{tab:stat}. (Color online.)}
\label{fig:clustCu}
\end{figure}

Fig.~\ref{fig:clustCu} shows copper-containing fragments of models 1,2, and 3.
These fragments illustrate typical structural motifs of the immediate
environments of the Cu atoms.
As seen in Table~\ref{tab:stat}, the coordination numbers of the Cu atoms
strongly correlate with the atomic charges, which indicates the covalent
nature of the bonding.
The ratio of the Mulliken charges (as defined in, e.g.,
Ref.~\cite{Simdyankin_2004}) of four- and two-fold coordinated Cu atoms
is about two, which suggests, along with the very nearly planar geometry of
the four-fold coordinated Cu atom in Fig.~\ref{fig:clustCu}(c), that such
atoms are in (II) and (I) oxidation states \cite{Greenwood_CoE}, respectively.
Three-fold coordinated Cu atoms (Fig.~\ref{fig:clustCu}(b)) may be viewed as
defective Cu(II) centers, also due to their nearly planar geometry and reduced
symmetry~--- the bond angles are approximately 90$^{\circ}$, 90$^{\circ}$, and
180$^{\circ}$.

Although the majority of Cu bonds are with S atoms (with a typical bond length
2.3-2.4~\AA), some Cu-As bonds (with lengths of about 2.3~\AA) are also
present in the models.
Invariably, one or both neighbors of the two-fold coordinated Cu(I) atoms were
found to be overcoordinated (three-fold coordinated S, e.g. S1 in
Fig.~\ref{fig:clustCu}(a), or four-fold coordinated As, see
Fig.~\ref{fig:clustCu}(d)), while the bonds of Cu(II) atoms are with normally
coordinated two-fold S and three-fold As atoms.

\begin{table}%[t]
\caption{Coordination numbers and Mulliken charges (in atomic units) of copper
atoms.}
\begin{tabular*}{8cm}{@{\extracolsep{\fill}}r*{2}{c}r}
 \hline \hline
\multirow{2}{*}{Model} & Number of & Coordination & Mulliken\\
                       & Cu atom   & number       & charge  \\ 
\hline
 1                  & 1 & 2 & 0.362 \\ 
\hline
\multirow{2}{*}{2}  & 1 & 4 & 1.050 \\
                    & 2 & 2 & 0.503 \\
\hline
\multirow{4}{*}{3}  & 1 & 3 & 0.826 \\
                    & 2 & 3 & 0.805 \\
                    & 3 & 2 & 0.360 \\
                    & 4 & 2 & 0.426 \\
\hline
\multirow{4}{*}{3a} & 1 & 2 & 0.400 \\
                    & 2 & 2 & 0.414 \\
                    & 3 & 2 & 0.420 \\
                    & 4 & 2 & 0.362 \\
\hline \hline
\end{tabular*}
\label{tab:stat}
\end{table}

\begin{figure} [t]
\centerline{(a)\includegraphics[width=8cm]{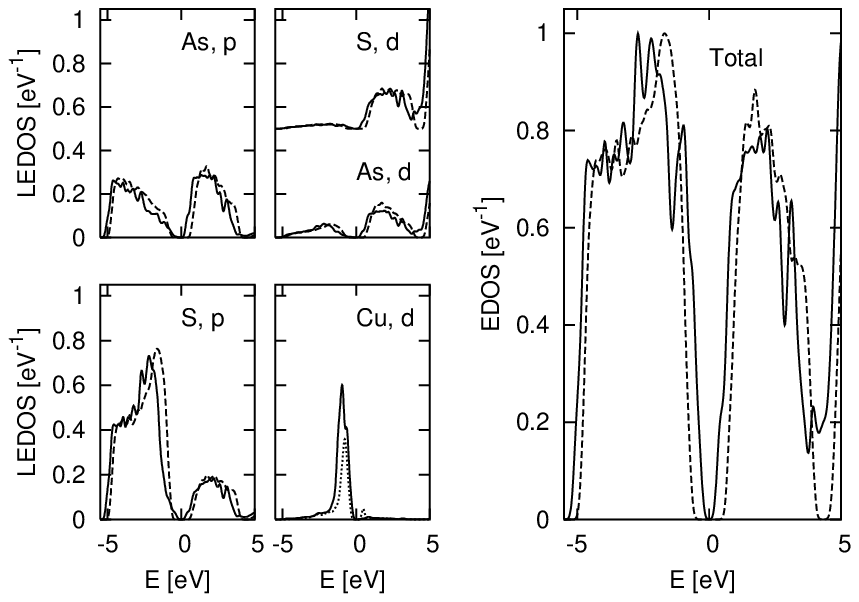}}
\centerline{(b)\includegraphics[width=8cm]{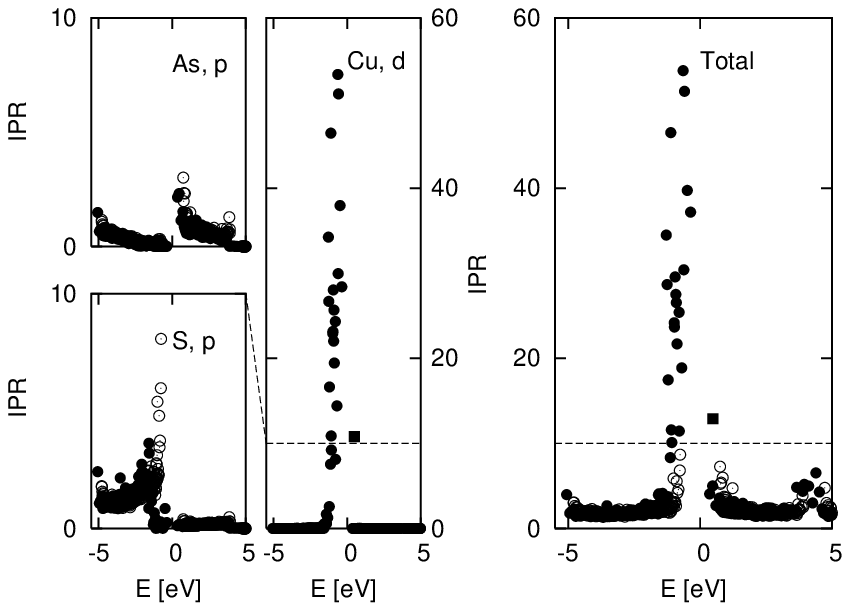}}
\caption{(a) Total and local EDOS for models 0 (dashed line) and 3a (solid
line) near the optical gap ($E=0$). The LEDOS for S, $d$ is shifted upwards by
0.5. The dotted line in the ``Cu, $d$'' panel corresponds to model 2. (b)
Total and projected inverse participation ratios for models 0 ($\odot$) and 3
({\LARGE $\bullet$}). The dashed lines emphasize that the vertical scales of
the ``As and S, $p$'' and ``Cu, $d$'' and ``Total'' axes are different. The
symbol ($\blacksquare$) in the ``Cu, $d$'' and ``Total'' panels corresponds to
model~2.}
\label{fig:edos_cu4}
\end{figure}

Fig.~\ref{fig:edos_cu4} shows the electronic densities of states (EDOS) and
inverse participation ratios (IPR) for models 0, 2, and 3a.
Mathematical definitions of both quantities, along with their local
(projected) variants, are given, e.g., in Ref.~\cite{Simdyankin_2004}.
The IPRs are normalized so that they are equal to unity for a totally
delocalized state (equally shared by all atoms) and to $N$, the number of
atoms in the model, for a state localized on only one atom.
The most obvious distinction between the EDOS's of models 0 and 3a is the
difference in the optical band gaps~--- the half-maximum gaps of models 0 and
3a are 1.95 eV and 1.34 eV, respectively, which accounts for the difference in
the optical band gaps of pure and copper-containing a-As$_2$S$_3$ observed
experimentally \cite{Liu_1990,Adriaenssens_2000}.
Inspection of the local EDOS (LEDOS) reveals that the main effect of the
addition of copper is to modify the top of the valence band. One can say that
the states due to sulfur lone-pair $p$ orbitals are pushed down in energy, and
the top of the valence band is now formed by highly localized Cu $d$ orbitals.
Copper atoms with any coordination contribute to the valence-band states,
while only three- and four-fold coordinated copper atoms produce additional
states at the bottom of the conduction band, as shown by the dotted line in
Fig.~\ref{fig:edos_cu4}(a) and solid squares in Fig.~\ref{fig:edos_cu4}(b).

\begin{figure} [t]
\centerline{\includegraphics[width=8cm]{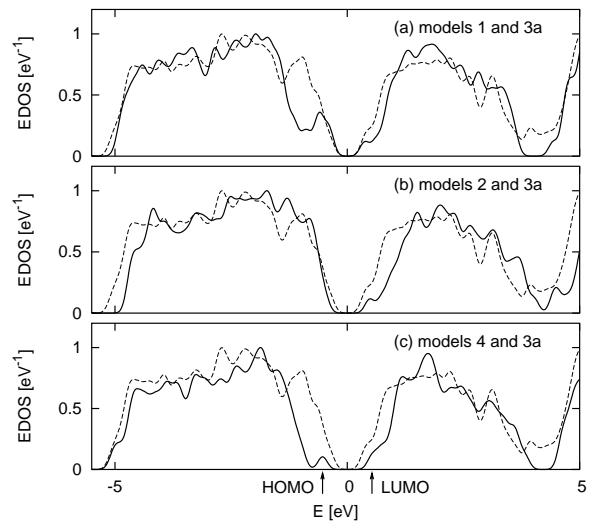}} 
\caption{EDOS near the optical gap for models 1 (a), 2 (b), and 4 (c) (solid
lines). The dashed line in all panels is the same and corresponds to model 3a
(same as the solid line in Fig.~\ref{fig:edos_cu4}(a)).  Arrows mark the HOMO
and LUMO energies in model 4.}
\label{fig:edoses}
\end{figure}

The above results show that the size of the optical band gap is determined by
the Cu atoms in copper-containing a-As$_2$S$_3$.
The Cu electronic states will mix with and mask any states due to native or
photoinduced defects in the amorphous network of As and S atoms positioned at
the same energies.
At low Cu concentration, a significant fraction of the volume in the material
is occupied by a copper-free As-S network, which should react to light in the
same way as would a pure binary alloy, as suggested by experimental results
\cite{Bolle_1998}.
It can be seen in Fig.~\ref{fig:edos_cu4}(b) that, although the S $p$ states
are located deeper in the valence band because of the Cu states, they are
still somewhat localized and qualitatively the localization character of S and
As $p$ states is preserved after addition of copper~--- the closer S~$p$
(As~$p$) states are to the top of the valence (bottom of the conduction) band,
the more localized they are.
Near-band-gap photoexcitations in this material can then be viewed as charge
transfers from the normally electronegative S atoms, where the state from
which an electron has been removed is predominantly localized, to the normally
electropositive As atoms, where the state to which the photoexcited electron
has been added is predominantly localized.

It has been shown in Ref.~\cite{Simdyankin_PRL_2005} that the excess of
negative and positive charge in the vicinity of electropositive As and
electronegative S atoms, respectively, produced by a photoexcitation can lead
to the formation of hypervalent [As$_4$]$^-$-[S$_3$]$^+$ defect pairs as an
alternative to radiative recombination of the photoinduced electron-hole
pairs.
Fig.~\ref{fig:edoses}(a,b) shows how the EDOS changes in the optical band-gap
region in the Cu-containing models with different copper contents, while
Fig.~\ref{fig:edoses}(c) compares the EDOS of copper-free model~4 (same as
model~2(III) in Ref.~\cite{Simdyankin_PRL_2005}), containing an
[As$_4$]$^-$-[S$_3$]$^+$ pair, with the EDOS of model 3a.
The highest occupied (HOMO) state in model 4 is localized along the chain of
three atoms forming the crossbar of the seesaw-like [As$_4$]$^-$ center,
similar to the S3-As1-S5 chain in Fig.~\ref{fig:clustCu}(a), and the lowest
unoccupied (LUMO) state is localized along an As-S bond, similar to the As2-S2
bond in Fig.~\ref{fig:clustCu}(a), where the S atom is an [S$_3$]$^+$ center.
The energies of these HOMO and LUMO states are marked in
Fig.~\ref{fig:edoses}(c), where it is seen that they are approximately equal
to the energies of the Cu-related states in Fig.~\ref{fig:edoses}(a) and (b).
The HOMO and LUMO energies in other models containing [As$_4$]$^-$-[S$_3$]$^+$
pairs are very close to those in model 4.

In model 1, the IPR of the LUMO state at the bottom of the conduction band is
largest in the region near the [S$_3$]$^+$ center (marked S2 in
Fig.~\ref{fig:clustCu}(a)), and the IPR of the seventh state counting from the
top of the valence band (HOMO-7) is highest for the singly-coordinated sulfur
atom (S5 in Fig.~\ref{fig:clustCu}(a)).
In this respect, the HOMO-7 state is similar to the HOMO state of copper-free
models.
Promoting one electron from HOMO-7 to LUMO exerts forces on the local
configuration shown in Fig.~\ref{fig:clustCu}(a) which tend to reduce the
distance between the atoms As1 and S5 (3.02~\AA~ in the ground state) and
elongate the bond S2-As2, precisely as in our time-dependent DFTB calculations
on a cluster containing an [As$_4$]$^-$-[S$_3$]$^+$ pair in
Ref.~\cite{Simdyankin_PRL_2005}.
Thus the As1,S2-5 group of atoms can be viewed as a precursor of a seesaw-like
[As$_4$]$^-$ center, and its presence provides a channel for electron-phonon
coupling that is expected to be large in these materials
\cite{Atta-Fynn_2004}.

%\section{Conclusions}

In summary, we have demonstrated that the main influence of small
concentrations of copper on the electronic structure and optical properties of
amorphous arsenic sulfide is to reduce the size of the optical band gap.
The photoinduced reduction of the size of the optical band gap in pure arsenic
chalcogenides, known as photodarkening, is of a similar magnitude and is not
expected to be discerned in the background of Cu-state-related electronic
photoexcitations.
This is true even for the lowest Cu concentration, i.e. 1.6\% (c.f. 1\%
\cite{Liu_1987}).
It is noted that Cu atoms are observed in typical Cu(I) and Cu(II)
configurations, and the four-fold coordinated Cu(II) centers are not
tetrahedral, but square planar.
This work also provides additional evidence that the hypervalent [As$_4$]$^-$
centers play a role in photoinduced phenomena in both Cu-free arsenic
chalcogenides, which is supported by the experimental observation
\cite{Zakery_1996} of a direct correlation of arsenic content with the
magnitude of photodarkening in these materials, and Cu-containing arsenic
chalcogenides.

%\section*{Acknowledgements}

S.I.S. is grateful to the EPSRC and Newton Trust for financial support.  We
thank the British Council and DAAD for provision of financial support.  

%\bibliographystyle{plain}
%\bibliographystyle{apsrev}
%\bibliography{archive}

\end{document}